# Miura-ori tube metamaterial with tunable dynamic property

Jie Liu[1,2*], Haifeng Ou[1], Rong Zeng[3], Wei Wei[1], Junfeng He[1], Jiaxi Zhou[3],

Wei Wei[2], Kai Long[4], Guilin Wen[1,2*], Yi Min Xie[5,6]

[1]Center for Research on Leading Technology of Special Equipment, School of Mechanical and Electric Engineering, Guangzhou University, Guangzhou 510006, P. R. China

[2]Center for Intelligent Equipment and Internet-connected Systems, Guangzhou University, Guangzhou 510006, P. R. China

[3]College of Mechanical and Vehicle Engineering, Hunan University, Changsha 410082, P.R. China

[4]State Key Laboratory for Alternate Electrical Power System with Renewable Energy Sources, North China Electric Power University, Beijing 102206, China

[5]Centre for Innovative Structures and Materials, School of Engineering, RMIT University, Melbourne 3001, Australia

[6]XIE Archi-Structure Design (Shanghai) Co., Ltd., Shanghai 200092, China

*Correspondence and requests for materials should be addressed to G.L.W. (email: glwen@gzhu.edu.cn) and J.L. (email: jliu@gzhu.edu.cn)

**Abstract:**

Structures and/or materials with engineered functionality, capable of achieving targeted mechanical responses reacting to changes in external excitation, have various potential engineering applications, e.g. aerospace, oceanographic engineering, soft robot, and several others. Yet tunable mechanical performance is normally realized through carefully designing the architecture of structures, which is usually porous, leading to the complexity of the fabrication of the structures even using the recently emerged 3D printing technique. In this study we show that origami technique can provide an alternative solution to achieving the aim by carefully stacking the classical Miura sheets into the Miura-ori tube metamaterial and tuning the geometric parameters of the origami metamaterial. By combining numerical and experimental methods, we have demonstrated that an extremely broad range of natural frequency and dynamic response of the metamaterial can be achieved. The proposed structure can be easily fabricated from a single thin sheet made of one material and simultaneously owns better mechanical properties than the Miura sheet.

**Keywords:** Miura sheets; Miura-ori tube metamaterial; tunable dynamic property

# 1. Introduction

Dynamic properties are of crucial significance and should be seriously considered when designing and developing engineering structures, e.g. aircraft, helicopter, missile, and many others [1,2]. In general, engineers attempt to maximize the fundamental frequency or the gap between the first two natural frequencies of one structure [3-12], with the aid of continuum topology optimization methods [3-22], to avoid the resonance which is normally critically harmful to the real-life structure .

Although these optimal designs have significantly improved the dynamic properties of one structure, they have two main drawbacks. The first one is these optimal designs are normally porous and very complicated [10], thereby leading to the difficulty in fabrication, even utilizing the recent emerged 3D printing technique; while the second one is the dynamic property of one structure is determined and can not be changed once designed and manufactured. It may become extremely non-optimal if the external dynamic stimuli changes. Hence, structures with tunable dynamic properties are excessively desired in engineering. 2D or 3D lattices can achieve tunable mechanical properties but also has fabrication problems [23-30]. Moreover, it generally has a relatively small tunable range. Is it possible to design a structure possessing a wide range of tunable dynamic properties and simultaneously easily fabricated? Origami technique may be one promising candidate.

Origami has gained ever increasing attention and has been employed to design various novel engineering structures from nanoscale [31-33] to large scale [34-38]. The goal of origami is to transform a flat square sheet of paper into an elaborate 3D structure through folding and sculpting techniques. It thus has fruitful merits such as easy to fabricate, avoiding the complex assembly, and normally lightweight. Substantial efforts have been devoted recently to design structures with tunable mechanical properties realized by origami techniques as well. Fuchi et al. [39] demonstrated that periodically decorating split-ring resonators on the origami surface can achieve tunable reflection and transmission characteristics. This is because origami technique can readily cause the geometry of the folded surface varied and thus lead to the change in distance between the rings generates a shift in resonance frequency, thereby resulting in the tunable mechanical properties. Previous studies showed that Poisson's ratio can be tuned from negative to positive with the folding angle changing by classical Miura-ori metamaterial [40] and Tachi-Miura polyhedron based reentrant 3D origami metamaterial [41]. By investigating the application potential of the origami technique in the thermodynamic field, Boatti et al. [42] elucidated that origami metamaterial can actualize an extremely broad range of thermal expansion coefficients by tuning the geometric parameters and elaborately arranging the facets and plates. Electromagnetic responses can also be dynamically tuned by origami metamaterial [43]. Recently, Jiang and his co-workers [44] created an origami-inspired metamaterial that holds tunable stiffness ability. It successfully overcomes the shortcomings of the existed deployable metamaterials that usually have unstable states.

However, to the authors' best knowledge, very few studies have been reported on origami metamaterials exhibit tunable dynamic properties [45-47]. Here, in this study, we have created a novel origami metamaterial by carefully tessellating two same Miura sheets and will show how the origami metamaterial can be tuned in terms of dynamic properties by altering the geometric parameters. Tunable first three natural frequencies are first investigated numerically and experimentally. Based on this, we numerically study the tunable dynamic displacement responses by altering the geometric parameters, i.e. the folding angle, the acute angle, the ratio of *a* to *b*, and the number of the unit cells.

The paper is organized as follows. After the Introduction, section 2 describes the geometric modelling of the proposed Miura-ori tube metamaterial. In section 3, finite element (FE) modelling and experimental setup for the natural frequency of the Miura-ori tube metamaterial are presented. Section 4 shows the results and discussions. The manuscript is closed with concluding remarks in section 5.

## 2. Geometrical modelling

In this section, the geometrical definition for the classical Miura sheet is first presented. On the basis of this, the way to construct the Miura-ori tube metamaterial will be shown.

*2.1. Miura sheet*

A Miura sheet cell is formed by four equivalent parallelograms, defined by a length $a$, a width $b$, and an acute angle $\beta$. Besides these three parameters, the dihedral angle (or folding angle), $\theta$, between two parallelograms is additionally needed to determine the geometric topology of the Miura sheet, as shown in Figure 1a. On the basis of the aforementioned four independent parameters, the dimensions $w$, $l$, $v$, and $h$, and the edge angles $\gamma$ and $\delta$ can be computed by using the following equations [37]

$$\cos\gamma = \sin^2\beta\cos\theta + \cos^2\beta \qquad (1)$$

$$\cos\delta = \frac{\sin^2\beta\cos^2(\theta/2) - \cos^2\beta}{\sin^2\beta\cos^2(\theta/2) + \cos^2\beta} \qquad (2)$$

$$w = 2b\sin(\gamma/2) \qquad (3)$$

$$h = a\cos(\delta/2) \qquad (4)$$

$$l = 2a\sin(\delta/2) \qquad (5)$$

$$v = b\sin(\gamma/2) \qquad (6)$$

To construct the Miura sheet, one can periodically arrange the unit cells in $x$ and $y$ directions. Figure 1b shows a typical Miura sheet with 4 by 1 unit cells. It should be noted that only one unit cell in $y$ direction will be used to create the Miura-ori tube metamaterial in this study, as discussed below.

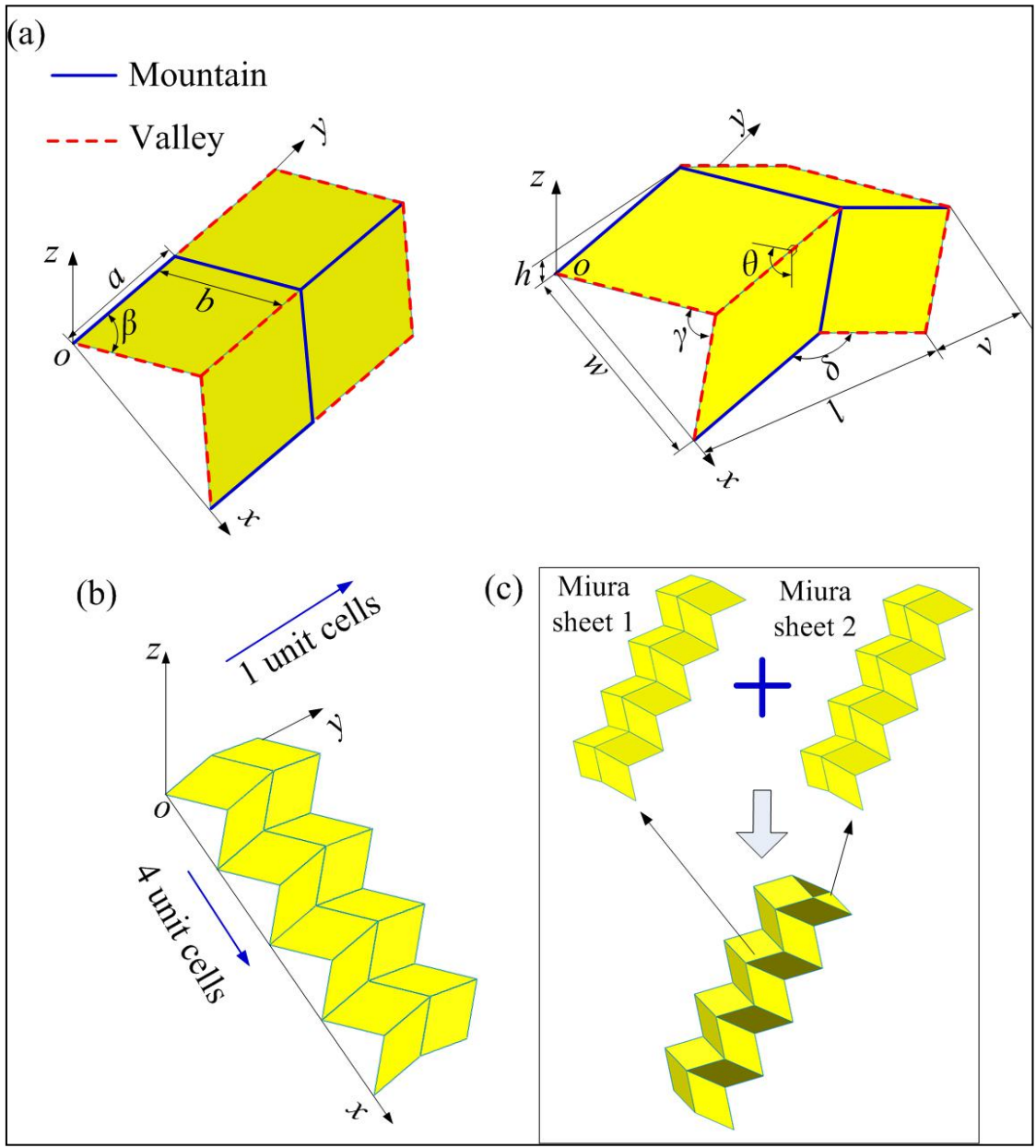

**Figure 1. Geometric modelling of the Miura-ori tube metamaterial**: (a) The unit cell of the Miura sheet, (b) a representative Miura sheet with 4 by 1 unit cells, and (c) tessellating two same Miura sheets to construct a Miura-ori tube metamaterial.

*2.2. Miura-ori tube metamaterial*

By carefully assembling two same Miura sheets at their open sides, one can easily construct a Miura-ori tube metamaterial. Figure 1c depicts a Miura-ori tube metamaterial comprised of two identical Miura sheets with 4 by 1 unit cells. It is worth underlining that once the constitute material is selected the geometric topology of the Miura-ori tube metamaterial is determined by five basic parameters, including four parameters that define the unit cell, namely, the length *a*, the width *b*, the acute angle *β*, and the folding angle *θ*, and the number of the unit cells in *x* direction,

*n*. Notice that the number of the unit cells in *y* direction remains to be 1. This means the targeted geometric topology of the Miura-ori tube metamaterial can be achieved by carefully adjusting these five parameters, which is significantly important for achieving the purpose we pursued in this study.

## 3. FE modelling and Experimental setup

### 3.1. FE modelling

To capture the natural frequency and other dynamic properties of the Miura-ori tube metamaterial metamateiral, numerical simulations are conducted using the commercial finite element code Abaqus/Linear perturbation. For calculating the natural frequency, Lanczos method implemented in Abaqus is used as the Eigensolver. Miura sheets 1 and 2 are rigidly connected at the edges of their open sides. In this work, the Miura-ori tube metamaterial is regarded as a cantilever with its left side fixed, which is meshed by using four-node shell elements with reduced integration (S4R). Brass (H62) is used as the constituted material in the Miura-ori tube metamaterial. To characterize the brass, tensile tests are performed by using a machine CMT5105 (Type: SUST) with a electronic extensometer (Type: YYU-12.5/25), as shown in Figure 2b. Tensile results indicate that the brass with a thickness $t$=0.2 mm can be characterized by a Young's modulus of $E$=103.6 GPa, yield stress $\sigma_y$=363.0 MPa, tensile strength $\sigma_u$=622.5 MPa, and elongation $\varepsilon_u$=18.8%. The Poisson's ratio is assumed as 0.35 and the density is found to be 8.9 g·cm$^{-3}$. The geometric parameters to determine the topology of Miura-ori tube metamaterial are: $a$=10 mm, $b$=10 mm, $\beta$=55°, $\theta$=130°, $n$=4.

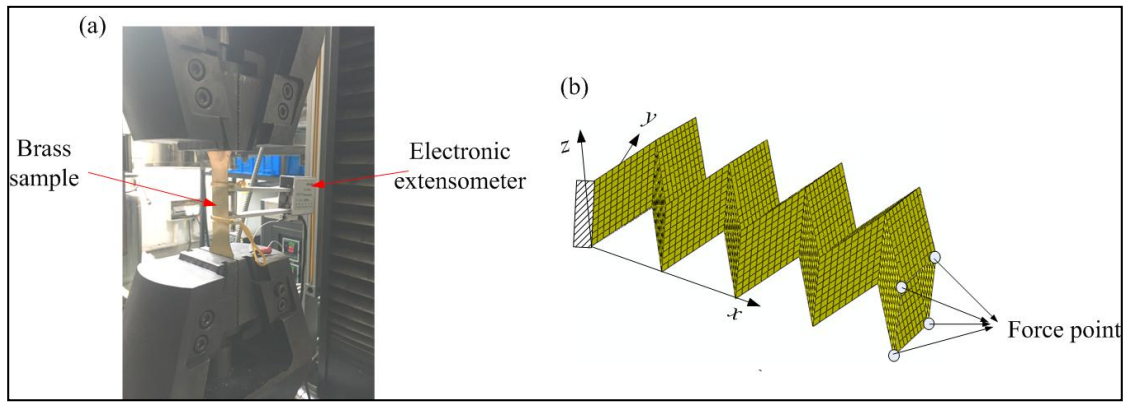

**Figure 2.** (a) Tensile test setup for obtaining the material properties of the brass sheet, and (b) FE modelling of the Miura-ori tube metamaterial cantilever.

For the dynamic response analysis, a harmonic load, $F(t)$, is equally distributed by four nodes on the right end of the structure (see Figure 2a). Here the expression of the harmonic load is assumed as, $F(t) = 2\sin(2\pi t)\,N$, $(0 \leq t \leq 5s)$, as shown in Figure 3a. We stress that the applied load is not restricted to the one we applied here. In this study, three cases are considered in terms of the position of the harmonic load applied, namely, on $xOy$ plane, $yOz$ plane, and $xOz$ plane, respectively (see Figures 3b-c, marked as Case 1, Case 2, and Case 3 for the ease of description in the following paragraphs).

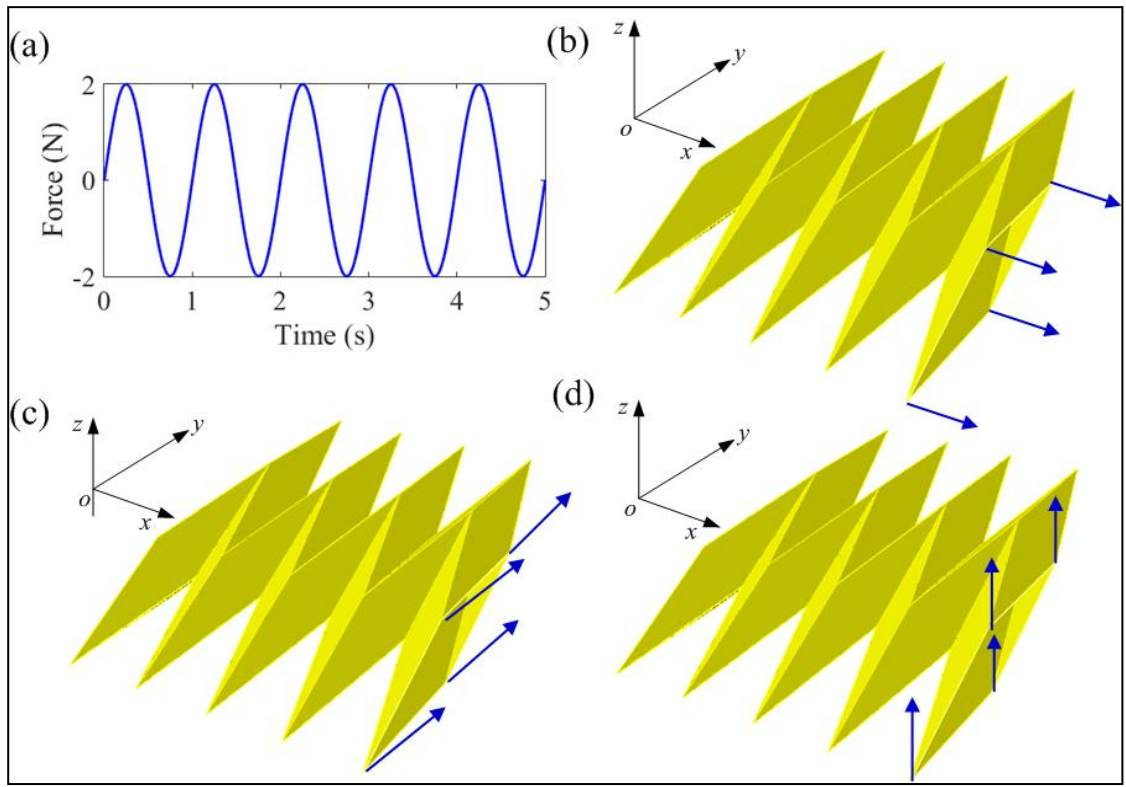

**Figure 3.** (a) Harmonic load, and (b)-(c) diagrams for the position of the applied load: $xOy$ plane, $yOz$ plane, and $xOz$ plane, respectively.

*3.2. Experimental setup*

*3.2. 1. Fabrication*

We use a five-step strategy to make the Miura-ori tube metamaterial, including cutting, moulding, stamping, trimming, and gluing, which is detailed as follows:

Step 1. Obtaining the flat sheet from the brass raw material by using a cutting machine. Note that carefully calculation is required to determine the material to make the Miura-ori tube metamaterial in this step.

Step 2. Manufacturing the male and female moulds for the Miura-ori tube metamaterial by using a commercial 3D printing machine Lite 450 with future 8000 photosensitive resin material.

Step 3. Putting the flat sheet on the middle of the male and female moulds and utilizing a punching machine (type: LY-WDQ20A4) to stamp the upper surface of the male mould. Miura sheet with a small amount of extra material is produced.

Step 4. Carefully trimming the extra material by using a hard scissors, particularly for the open sides of the Miura sheets.

Step 5. Repeating Steps 1-4 twice, and getting two approximately two same Miura sheets. Gluing the two Miura sheet at their open sides by using a commercial glue ergo 1690 to form the final Miura-ori tube metamaterial.

Figure 4a depicts three kinds of moulds fabricated with the folding angle as 90, 100, and 130 degree, respectively. Figure 4b shows a typical manufacturing process for the Miura-ori tube metamaterial.

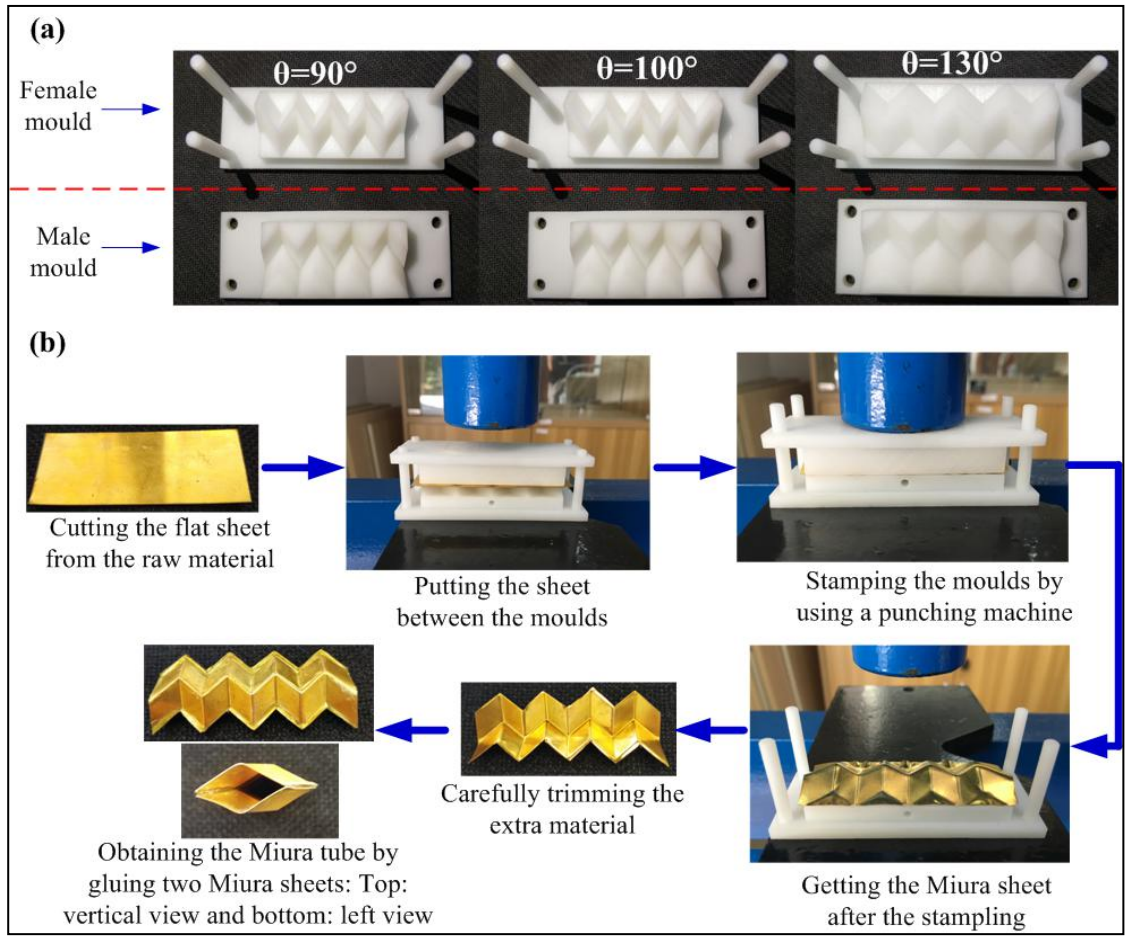

**Figure 4. Fabrication of the Miura-ori tube metamaterial.** (a) Moulds for creating the Miura sheets and (b) a representative manufacturing process.

*3.2. 2. Vibration testing*

Excitation method is employed to capture the first three natural frequencies of the Miura-ori tube metamaterial. As shown in Figure 5, one end of the test sample is clamped on a copper block which is fixed on a modal shaker (Type: JZK-40) and a laser triangulation sensor (Type: 15-28VDC/2.8W) is installed above the test sample. Three kinds of vibration excitation are generated by the modal shaker, namely, white noise, sinusoidal excitation with frequency band 200~2500 Hz, and sinusoidal excitation with frequency band 200~3000 Hz. The modal shaker causes the vibration of the test sample whose vibration signals are perceived by the laser triangulation sensor (collecting 10240 sample points per second for a total of 0.8 second). The output response signal is gained by using a data collector (Type: AVANT Lite MI–6008) and is amplified by a power amplifier. Finally, the output time domain signal is analyzed by frequency domain in Matlab, and the frequencies corresponding to the first, second and third order vibration modes are respectively obtained.

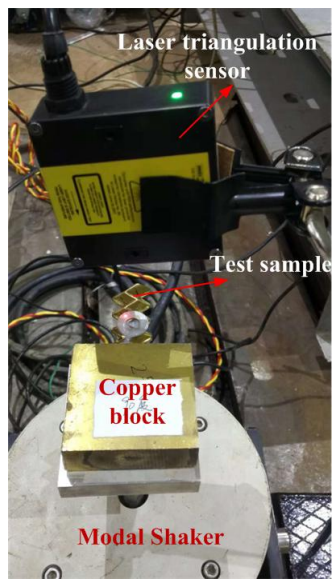

**Figure 5. The vibrating testing setup.**

## 4. Results and discussions

*4.1 Normalized natural frequency*

4.1.1 Comparison study

In this study, the first three normalized natural frequencies (NNF) are calculated, as $\bar{\omega}_i = \omega_i / \omega_i^{\theta=90°}$, $i = 1, 2, 3$. Here $\omega_i$ denote the natural frequencies of the Miura-ori tube metamaterial and $\omega_i^{\theta=90°}$ refer to the natural frequencies for the benchmark Miura-ori tube metamaterial when the folding angle, $\theta$, equals to 90°. From Figure 6, we numerically show that the NNFs, $\bar{\omega}_1$, $\bar{\omega}_2$ and $\bar{\omega}_3$ could be tuned in an extremely broad range by simply altering the folding angle from 50 deg to 130 deg, with their values as 0.73-1, 0.93-1.54, and 0.62-2.04, respectively. In addition, Miura-ori tube metamaterials with the folding angle as 90 deg, 100 deg, and 130 deg are fabricated and their first three natural frequencies are tested and summarized in Table 1. For ease of description, tests with white noise, sinusoidal excitation with frequency band 200~2500 Hz, and sinusoidal excitation with frequency band 200~3000 Hz are marked as Test 1, Test 2, and Test 3, respectively. The corresponding NNFs of the tests are also presented in Figure 6. It should be noticed that the Miura-ori tube metamaterial is very short mainly considering its manufacture. Thus it is not an easy task to precisely capture its natural frequencies. In this study, the NNFs obtained from the vibrating tests are used to qualitatively predict the natural frequencies of the Miura-ori tube metamaterial. It can be found from Figure 6 that, there is a little bit difference between the experimental and numerical results for $\bar{\omega}_1$ when the folding angle equals to 130 deg; while relatively large difference are found when the folding angle equals to 100 deg. For $\bar{\omega}_2$, the results from the tests are all near the sides of the simulation curve. However, there are relatively big deviation between the test results and the simulation results, which may be because of the distinction of the boundary conditions of the Miura-ori tube metamaterial between the experimental and numerical models. However, considering that we mainly focus on the tunable trend instead of its absolute exact value, the FE modelling therefore can be verified and employed to investigate the tunable dynamic properties of the Miura-ori tube metamaterial.

**Table 1** Results from vibration tests when the folding angle is 90°, 100°, and 130°

|  | $\omega_1$ (Hz) | $\omega_2$ (Hz) | $\omega_3$ (Hz) |
|---|---|---|---|
| $\theta=90°$ |  |  |  |
| Test 1 | 480 | 680 | 1850 |
| Test 2 | 450 | 650 | 1780 |
| Test 3 | 450 | 650 | 1780 |
| $\theta=100°$ |  |  |  |
| Test 1 | 800 | 900 | 1800 |
| Test 2 | 750 | 880 | 1820 |
| Test 3 | 750 | 880 | 1800 |
| $\theta=130°$ |  |  |  |
| Test 1 | 400 | 880 | 1850 |
| Test 2 | 400 | 880 | 1850 |
| Test 3 | 400 | 880 | 1850 |

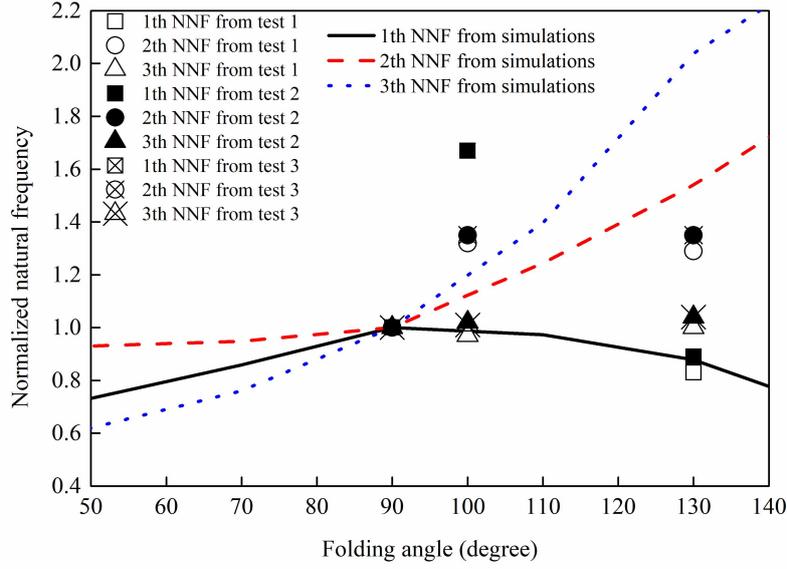

**Figure 6.** Experimental and numerical results.

4.1.2 Effects of the acute angle, $\beta$

To investigate the influence of the acute angle, $\beta$, on NNF, numerical simulations are performed by altering its value from 20° to 80° at interval of 30°. Two situations are considered in terms of the ration of $a$ to $b$. By fixing the value of $a$ as 10 mm, $a/b=1$ and $a/b=2$ are considered, which are depicted in Figure 7(a) and (b), respectively, revealing that the acute angle, $\beta$, significantly affect the NNF in both cases. Specifically, $\bar{\omega}_1$, $\bar{\omega}_2$, and $\bar{\omega}_3$ are all gradually increased by tuning the folding angle, $\theta$, from 50° to 130° when $\beta=80°$; for $\beta=50°$, interestingly, $\bar{\omega}_1$ slightly becomes larger first and then slowly tends to be smaller; the trends of NNF are extremely complex for $\beta=20°$, such that, $\bar{\omega}_1$ and $\bar{\omega}_2$ become smaller and bigger as the folding angle increases, respectively, but $\bar{\omega}_3$ increases first and then decreases. Similar phenomena are discovered when $a/b=2$ except for $\beta=20°$ that $\bar{\omega}_2$ gets larger very slightly first and then becomes smaller. It can be summarized that NNF can be tuned larger by diminishing the value of $\beta$. When $\beta=20°$, the maximum tuning range for $\bar{\omega}_1$, $\bar{\omega}_2$, and $\bar{\omega}_3$ are 1-1.9, 2.9-4.2, 2.9-3.9, ($a/b=1$), and 5.1-9.6, 10.5-13.1, 8.2-10.4, ($a/b=2$) respectively, thereby demonstrating that NNF can be tuned in an extremely broad range by adjusting $\theta$ and $a/b$.

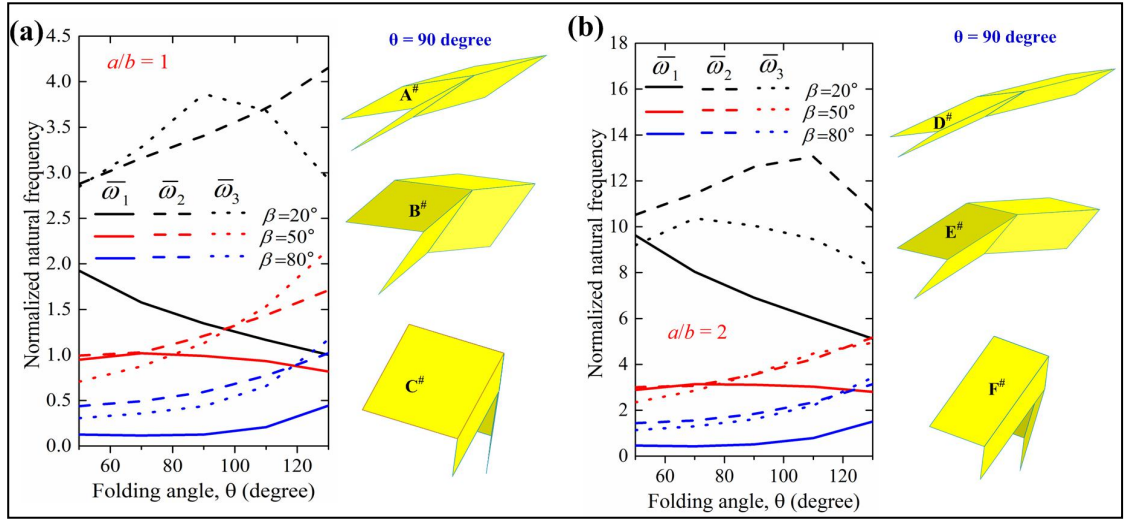

**Figure 7. Effects of $\beta$ on the normalized natural frequency when the folding angle, $\theta$, changes from 50 to 130 degree.** (a) $a/b$=1 (The right panel shows typical unit cells of the benchmark Miura-ori tube metamaterial when $\beta$ equals to 20°, 50°, and 80°, marked by $A^\#$, $B^\#$, and $C^\#$, respectively); (b) $a/b$=2 (The right panel presents typical unit cells of the benchmark Miura-ori tube metamaterial when $\beta$ equals to 20°, 50°, and 80°, marked by $D^\#$, $E^\#$, and $F^\#$, respectively).

*4.2 Tuning dynamic responses*

4.2.1 Altering the folding angle, $\theta$

To study how the dynamic displacement responses can be tuned by changing the folding angle, $\theta$, FE simulations are performed by altering the value of $\theta$ from 50° to 130°. To purely investigate the influence of the folding angle, we keep the following parameters constant such that $a$=10 mm, $n$=4, $a/b$=1, and $\beta$=55°. Figure 8a-c show the dynamic displacement responses versus time in $x$ direction, $y$ direction, and $z$ direction correspond to Case 1, Case 2, and Case 3 presented in Figure 3b-d, respectively. Changing the value of the folding angle results in the apparent effects on dynamic displacement response. For Case 1 and Case 2, the larger of the folding angle, the smaller of the amplitude of the vibration, but it is the opposite for Case 3. By comparing the maximum vibration amplitude, it can be found that in-plane stiffness (Case 1) is greater than the other two out-of-plane stiffness (Case 1 and Case 2), which is easy to understand. When the folding angle is 50°, the maximum vibration amplitudes for Case 2 and Case 3 are approximately 0.12 mm and 0.11 mm, respectively, revealing that these two out-of-plane stiffness are almost the same. However, by increasing the folding angle, i.e. when $\theta$ equals to 130°, the maximum vibration amplitudes become roundly 0.04 mm and 0.20 mm, respectively. It means that the Miura-ori tube metamaterial gets stiffer out of $xOz$ plane than that of out of $xOy$ plane. Thus, it can be concluded that we may design the Miura-ori tube metamaterial with the stiffness we expected by simply adjusting the value of the folding angle.

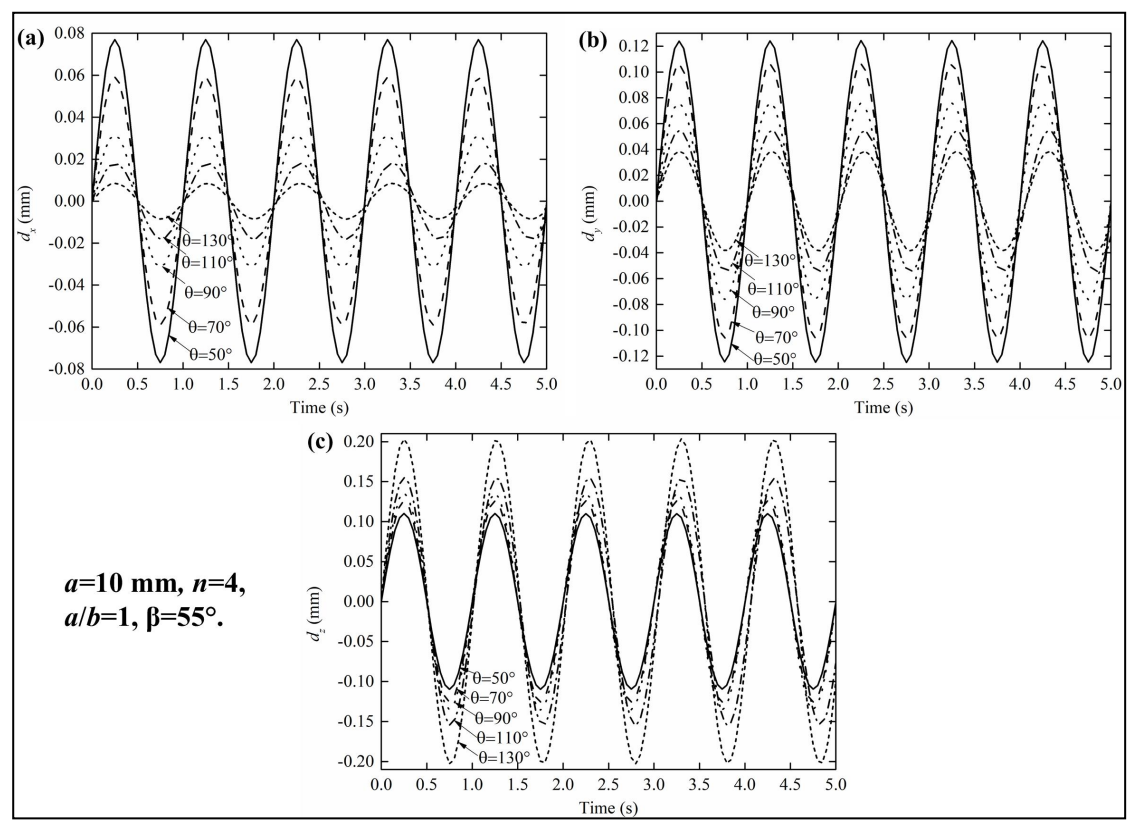

**Figure 8. Dynamic displacement responses versus time by altering the folding angle from 50°
to 130°.** (a) $d_x$, dynamic displacement in $x$ direction by applying the harmonic load in $xOy$ plane;
(b) $d_y$, dynamic displacement in $y$ direction by applying the harmonic load in $yOz$ plane; (c) $d_y$,
dynamic displacement in $y$ direction by applying the harmonic load in $xOz$ plane.

4.2.2 Altering the acute angle, $β$

By keeping the values of $a$, $n$, $a/b$, and $θ$ fixed, as 10 mm, 4, 1, and 50°, respectively, the dynamic displacement response can be tuned in a broad range via purely modifying the value of the acute angle, $β$, from 25° to 70°, as presented in Figure 9. For case 1, it can be clearly found that the maximum dynamic displacement response can be tuned from 0.019 mm to 0.273 mm; in other words, the maximum dynamic displacement response can be reduced approximately 93% by simply altering the acute angle $β$, from 70° to 25° (see Figure 9a). This is of great significance if one wants to alleviating the vibration (reducing the vibration amplitude) of one structure, but does not want to design structure with complex configurations which is usually hard to fabricate. It can be also found that the increasing rate of the dynamic displacement response becomes larger as gradually enlarges the the acute angle in the same size, namely, 15°, demonstrating that the in-plane stiffness of the Muira-ori tube metamaterial is rapidly dropped. This phenomenon is more prominent for Case 2, as shown in Figure 9b. No significant changes happen to the dynamic displacement response when altering the acute angle from 55° to 25°, but it excessively rockets up when $\beta$ equals 70°, revealing that the stiffness out of $xOz$ plane degenerates to be very small when $\beta$ near or larger than 70°. For Case 3, as shown in Figure 9c, it seems that the dynamic displacement response is not sufficiently tuned as one modifying the value of $\beta$; in other words, the dynamic displacement response is insensitive to $\beta$. Moreover, there is negligible difference in the dynamic displacement response when $\beta$ is 40° and 55°, meaning that one can find two Miura-ori tube metamaterial owning almost the same stiffness out of $xOy$ plane, which may be great useful for the real-life structure when its stiffness is specified but the space usage is limited.

To further investigate the generality of the tunable dynamic property by changing $\beta$, two more situations with the folding angle as 90° and 130° are considered as shown in Figure 10 and 11, respectively. Comparing these results with those in Figure 9, similar tunable dynamic property can be obviously observed. When the folding angle is 90° (see Figure 10), interestingly, two couples of Miura-ori tube metamaterial respectively have approximately the same stiffness out of $xOy$ plane, i.e. the metamaterial with $\beta$ = 25° and $\beta$ = 40°, and the metamaterial with $\beta$ = 55° and $\beta$ = 70°. However, this phenomenon disappears when the folding angle equals 130° (see Figure 11). Hence, again, it can be concluded that it is possible to obtain the Miura-ori tube metamaterial with dynamic property we anticipated by carefully adjusting value of the acute angle.

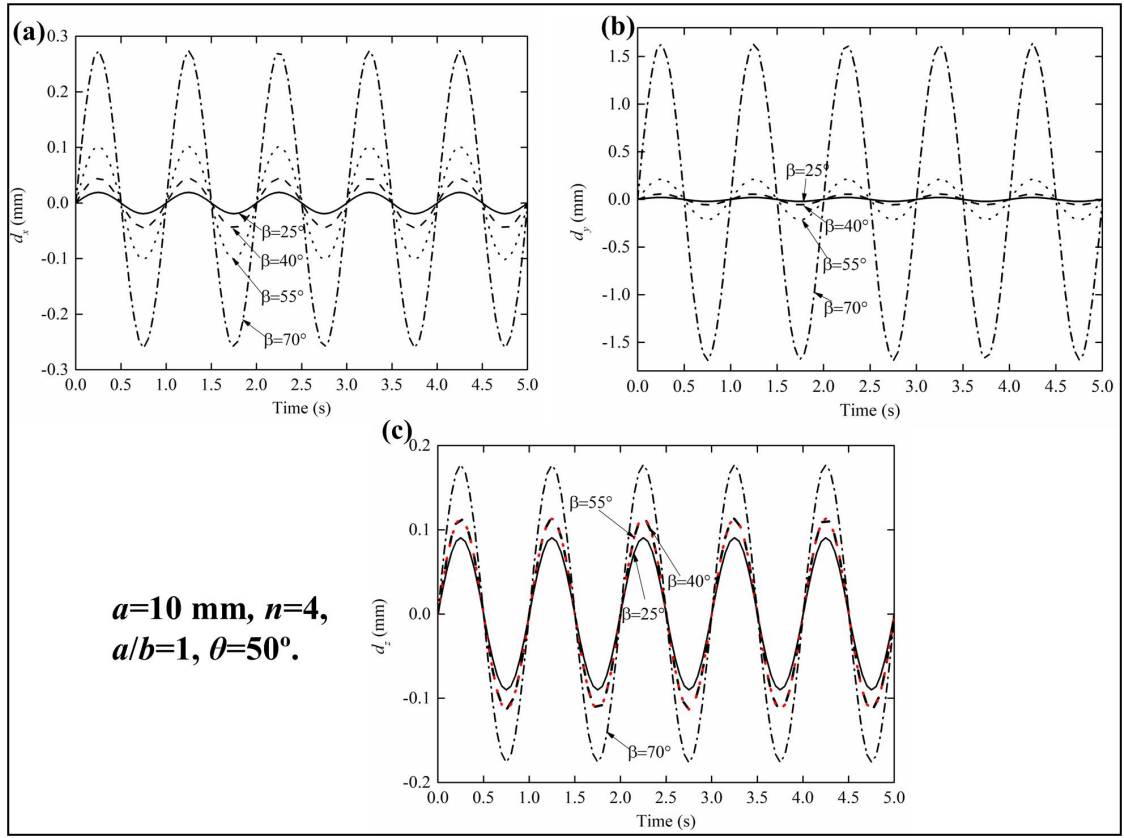

**Figure 9.** Dynamic displacement responses versus time by altering the acute angle from 25° to 70° when the folding angle equals to 50°. (a) $d_x$, dynamic displacement in $x$ direction by applying the harmonic load in $xOy$ plane; (b) $d_y$, dynamic displacement in $y$ direction by applying the harmonic load in $yOz$ plane; (c) $d_y$, dynamic displacement in $y$ direction by applying the harmonic load in $xOz$ plane.

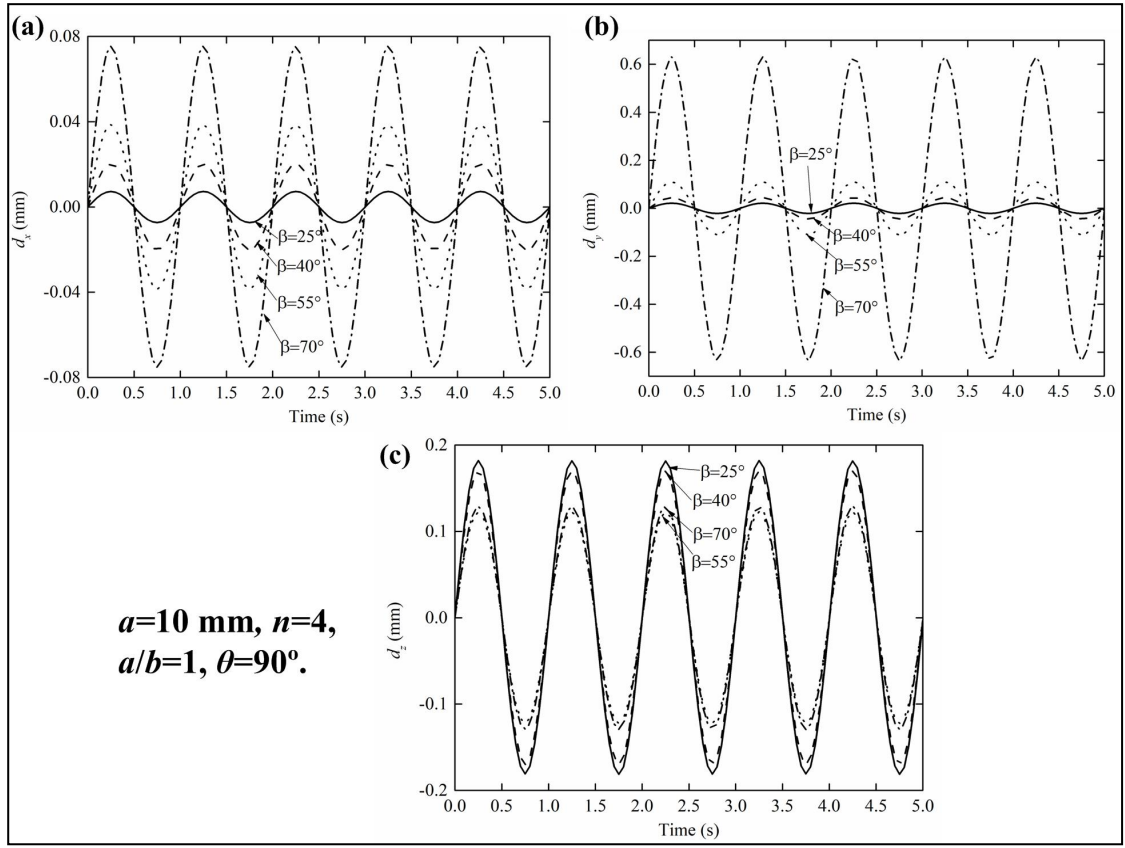

**Figure 10. Dynamic displacement responses versus time by altering the acute angle from 25°  to 70° when the folding angle equals to 90°.** (a) $d_x$, dynamic displacement in $x$ direction by applying the harmonic load in $xOy$ plane; (b) $d_y$, dynamic displacement in $y$ direction by applying the harmonic load in $yOz$ plane; (c) $d_y$, dynamic displacement in $y$ direction by applying the harmonic load in $xOz$ plane.

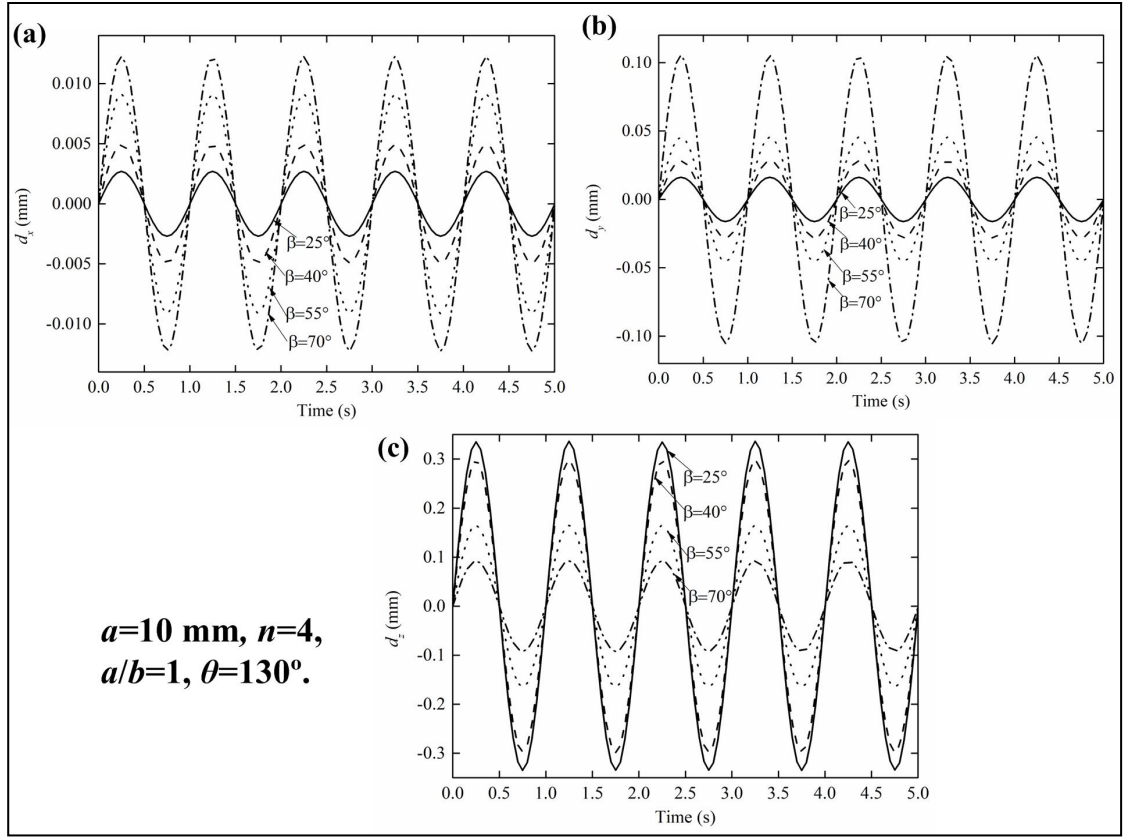

**Figure 11. Dynamic displacement responses versus time by altering the acute angle from 25°
to 70° when the folding angle equals to 130°.** (a) $d_x$, dynamic displacement in $x$ direction by
applying the harmonic load in $xOy$ plane; (b) $d_y$, dynamic displacement in $y$ direction by applying
the harmonic load in $yOz$ plane; (c) $d_y$, dynamic displacement in $y$ direction by applying the
harmonic load in $xOz$ plane.

4.2.3 Altering the ration of $a$ to $b$

To study how the dynamic displacement response can be tuned by altering the ration of $a$ to $b$, we keep the values of the following parameters unchanged, as $a$=10 mm, $n$=4, $β$=55°, and $θ$=130°, and conduct the FE analysis. For Case 1, $d_x$ has not varied much as depicted in Figure 12a, meaning that the in-plane stiffness is insensitive to the fluctuation of the ration of $a$ to $b$. However, $d_y$ and $d_z$ can be readily tuned by altering the ration of $a$ to $b$; specifically, simply modifying the ratio of $a$ to $b$ from 1.0 to 2.2 results in the tune of the maximum value of $d_y$ from 0.046 mm to 0.424 mm, altering nearly 10 times, and the maximum value of $d_z$ from 0.165 mm to 0.768 mm, changing roundly 5 times. It can be evidenced that dynamic displacement response of the Miura-ori tube metamaterial can be tuned in a wide range by simple altering the ration of $a$ to $b$. Moreover, comparing Figure 12b and Figure 12c, it can also be found that the stiffness of out of $xoz$ plane is almost two times larger than that of out of *xoy* plane.

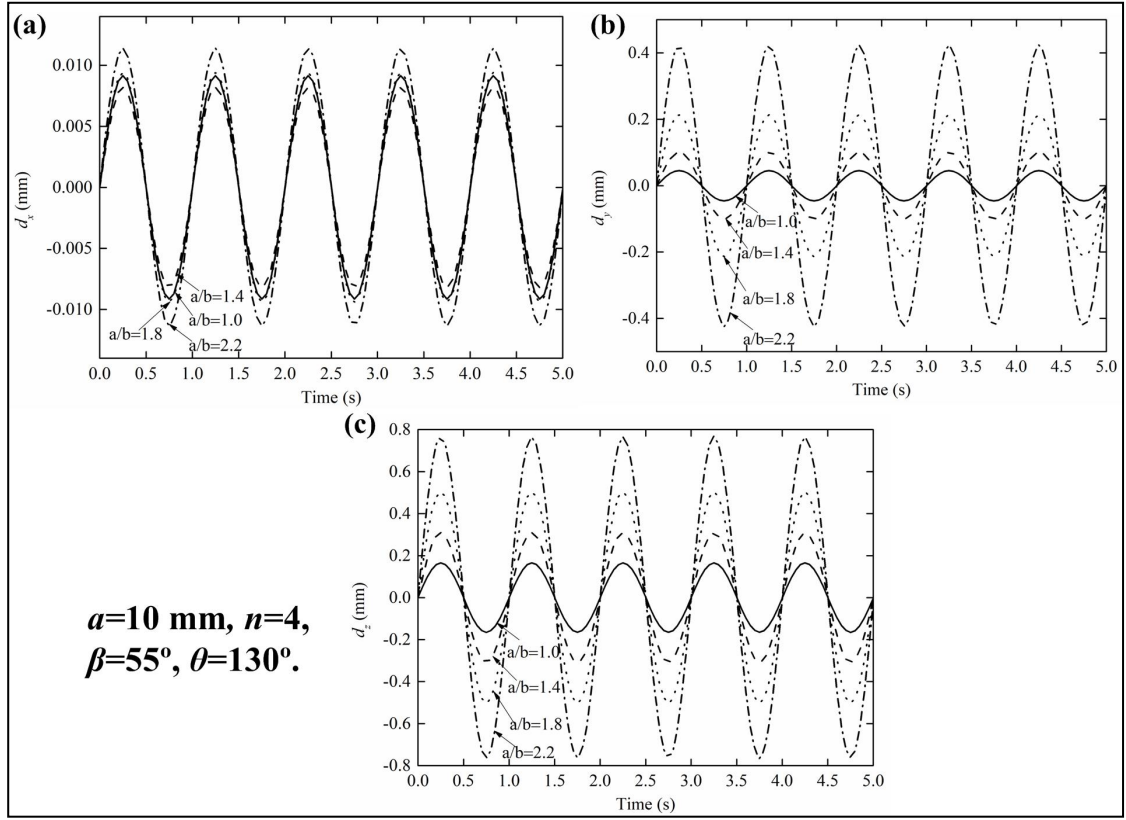

**Figure 12. Dynamic displacement responses versus time by altering the ration of *a* to *b* from 1.0 to 2.2** . (a) $d_x$, dynamic displacement in *x* direction by applying the harmonic load in *xOy* plane; (b) $d_y$, dynamic displacement in *y* direction by applying the harmonic load in *yOz* plane; (c) $d_y$, dynamic displacement in *y* direction by applying the harmonic load in *xOz* plane.

4.2.4 Altering the number of the unit cells, *n*

Besides the parameters discussed previously, the number of the unit cells, *n*, can significantly determine the tunability of the dynamic displacement response of the Miura-ori tube metamaterial, which will be presented in this section. It should be noticed that the material for constructing the Miura-ori tube metamaterial increases as the number of the unit cells becomes large, which differs from those discussed aforementioned, but it has no significant influence on what we aim to study. By enlarging the number of the unit cells from 4 to 10, the dynamic displacement response can be apparently tuned in all the three cases, namely, Case 1, Case 2, and Case 3, corresponding to Figure 13a-c, respectively. In particular, for Case 3, the maximum value of $d_z$ can be tuned in a extremely wide range, namely, from 0.17 mm to 2.87 mm, evidently demonstrating that the dynamic displacement response can be expediently tuned by altering the number of the unit cells.

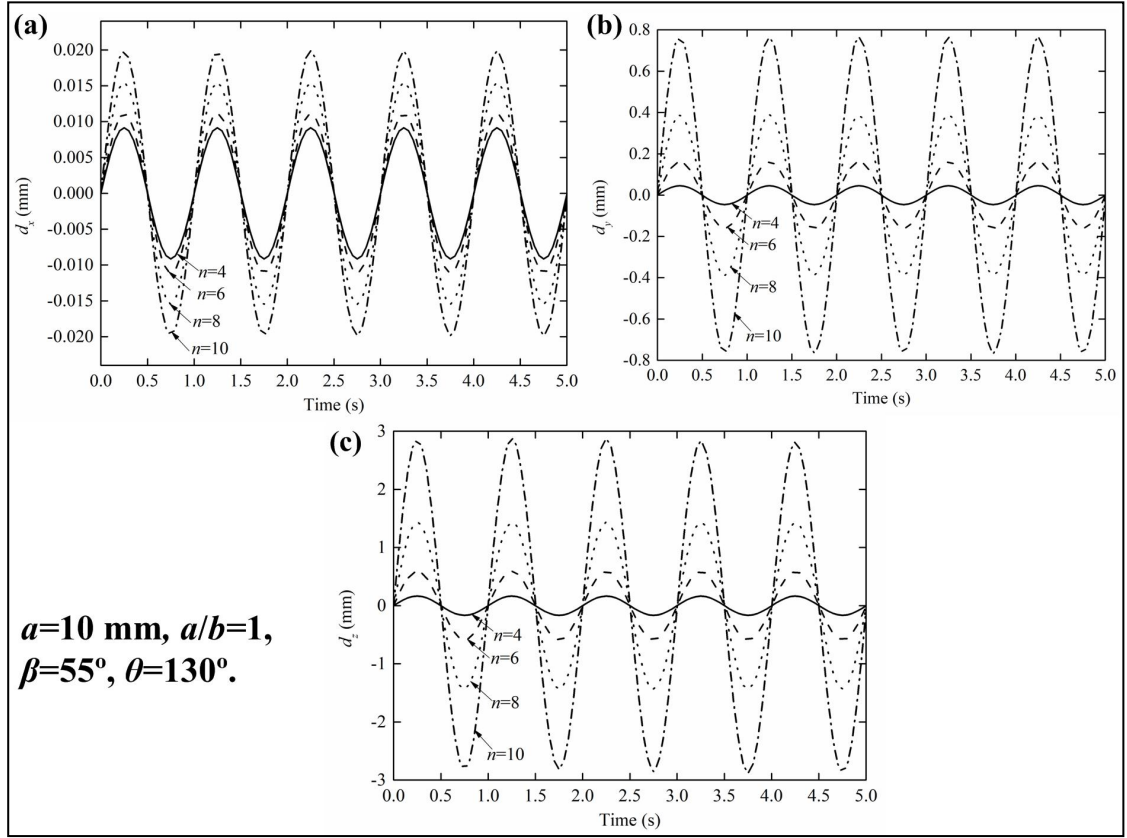

**Figure 13. Dynamic displacement responses versus time by altering the number of the unit cells from 4 to 10**. (a) $d_x$, dynamic displacement in $x$ direction by applying the harmonic load in $xOy$ plane; (b) $d_y$, dynamic displacement in $y$ direction by applying the harmonic load in $yOz$ plane; (c) $d_y$, dynamic displacement in $y$ direction by applying the harmonic load in $xOz$ plane.

*4.2.5 Theoretical understanding*

In this section, we will show, from the theoretical perspective, why the dynamic property of the Miura-ori tube metamaterial can be tuned.

The dynamic model for a structural system subjected to a harmonic load can be expressed as [48-50],

$$m\ddot{x} + c\dot{x} + kx = P_0 \cos \omega t \tag{7}$$

where $m$, $c$, and $k$ are the mass, linear damping coefficient, and stiffness, respectively. $P_0$ and $\omega$ represent the amplitude and the frequency of the harmonic load, respectively. $x$, $\dot{x}$ and $\ddot{x}$ refer to displacement, velocity, and acceleration, respectively.

Here, the damping of the system is neglected. Once the flat thin plate for constructing the Miur-ori metamaterial is determined, its mass is kept constant. In addition, the harmonic load is normally predefined. From Eq. (7), it can be seen that the NNFs and the dynamic response of the Miura-ori tube metamaterial is purely dominated by its stiffness. In other words, the NNFs and the dynamic response can be easily tuned if the stiffness can be conveniently modified. Fortunately, the stiffness can be simply altered by changing the geometric topology of the Miura-ori tube metamaterial. Therefore, it can be theoretically elucidated that the dynamic response can be tuned by adjusting the folding angle, the acute angle, the ration of $a$ to $b$, and the number of the unit cells.

## 5. Concluding remarks

In this study, we have proposed a novel Miura-ori tube metamaterial by carefully tessellating two same Miura sheets. By mean of numerical simulations and experiments, it is demonstrated that the first three NNFs of the Miura-ori tube metamaterial can be tuned in an extremely wide range by changing the folding angle and the ratio of $a$ to $b$. We then numerically investigate the dynamic displacement response for three cases in terms of the position of the harmonic load applied, manifesting that it can be tuned in a wide range, particularly for Case 2 and Case 3. For instance, the maximum value of $d_z$ can be tuned from 0.17 mm to 2.87 mm by enlarging the number of the unit cells from 4 to 10 for Case 3. The reason why the proposed Miura-ori tube metamaterial can be tuned dynamically has been also qualitatively explained from the theoretical perspective. These results open a new avenue toward lightweight and reconfigurable metamaterials with simultaneously engineered tunable dynamic properties. Moreover, unprecedented opportunities for lightweight structures to meet the demands with extremely wide range of tunable dynamic properties will be provided when multiple materials are used to constitute the Miura-based tube metamaterial.

Further research will focus on investigating the influences of the damping [50] on the tunable dynamic properties. Moreover, we will attempt to establish the dynamic model of the Miura-ori tube metamaterial by using the spring-mass-damping system [51] and derive the formulations of NNFs and dynamic displacement responses.

**Acknowledgements**

This research was financially supported by the Key Program of National Natural Science Foundation of China (No. 11832009), the National Natural Science Foundation of China (No.11672104), the Chair Professor of Lotus Scholars Program in Hunan province (No.XJT2015408), and the National Natural Science Foundation of Beijing (No. 2182067). Meanwhile, the authors are very grateful to the reviewers for their valuable comments and suggestions.